\documentclass[conference]{IEEEtran}
\IEEEoverridecommandlockouts
\usepackage{cite}
\usepackage{amsmath,amssymb,amsfonts}
\usepackage{algorithmic}
\usepackage{graphicx}
\usepackage{textcomp}
\usepackage{xcolor}
\usepackage{listings}
\usepackage{tipa}
\usepackage{footmisc}

\definecolor{codegreen}{rgb}{0,0.6,0}
\definecolor{codegray}{rgb}{0.5,0.5,0.5}
\definecolor{codepurple}{rgb}{0.24,0,0.41}
\definecolor{backcolour}{rgb}{0.96,0.96,0.96}

\usepackage{lineno}

\lstdefinestyle{cstyle}{
    language=C,
    frame=lines,
    commentstyle=\color{codegreen},
    numberstyle=\tiny\color{codegray},
    stringstyle=\color{magenta},
    basicstyle=\ttfamily\footnotesize,
    breakatwhitespace=false,         
    breaklines=true,                 
    captionpos=b,                    
    keepspaces=true,                 
    numbers=left,                    
    numbersep=5pt,                  
    showspaces=false,                
    showstringspaces=false,
    showtabs=false,                  
    tabsize=2
}

\def\BibTeX{{\rm B\kern-.05em{\sc i\kern-.025em b}\kern-.08em
    T\kern-.1667em\lower.7ex\hbox{E}\kern-.125emX}}
\begin{document}

\title{Benchmarking micro-core architectures for detecting disasters at the edge}

\author{\IEEEauthorblockN{Maurice Jamieson}
\IEEEauthorblockA{\textit{EPCC} \\
\textit{University of Edinburgh}\\
Bayes Centre, 47 Potterrow, Edinburgh \\
maurice.jamieson@ed.ac.uk}
\and
\IEEEauthorblockN{Nick Brown}
\IEEEauthorblockA{\textit{EPCC} \\
\textit{University of Edinburgh}\\
Bayes Centre, 47 Potterrow, Edinburgh}
}

\maketitle

\begin{abstract}
Leveraging real-time data to detect disasters such as wildfires, extreme weather, earthquakes, tsunamis, human health emergencies, or global diseases is an important opportunity. However, much of this data is generated in the field and the volumes involved mean that it is impractical for transmission back to a central data-centre for processing. Instead, edge devices are required to generate insights from sensor data streaming in, but an important question given the severe performance and power constraints that these must operate under is that of the most suitable CPU architecture. One class of device that we believe has a significant role to play here is that of micro-cores, which combine many simple low-power cores in a single chip. However, there are many to choose from, and an important question is which is most suited to what situation.
This paper presents the Eithne framework, designed to simplify benchmarking of micro-core architectures. Three benchmarks, LINPACK, DFT and FFT, have been implemented atop of this framework and we use these to explore the key characteristics and concerns of common micro-core designs within the context of operating on the edge for disaster detection. The result of this work is an extensible framework that the community can use help develop and test these devices in the future.
\end{abstract}


\section{Introduction}
The ability to detect and track the unfolding of disasters is often enabled by the Internet of Things (IoT), where devices sit out \emph{on the edge}, and are used to monitor a series of sensors. Whether it be detecting wildfires, earthquakes, extreme weather, or excessive pollution, these systems must work reliably within challenging tolerances. One of the most challenging aspects is that, because they are often battery powered, the IoT devices must work efficiently yet draw minimal amounts of power. Desktop and server processor designs, such as the Intel i7 and Xeon, have significant power draw requirements and as such are completely inappropriate for such workloads.

Micro-core architectures look to deliver high performance whilst addressing power consumption issues by implementing large numbers of simple cores running in parallel on a single chip. There have been some successes in deploying such architectures in the data centre for HPC workloads, such as the Sunway SW26010 of the Sunway TaihuLight  (\#3 June 2019 Top500 \cite{June2019TOP5002019}) and the 2048 core PEZY-SC2 of the Shoubu system B (\#1 June 2019 Green500 \cite{June2019GREEN5002019}). However, more generally micro-cores deliver significant benefits out in the field \cite{labowski2016implementing} running embedded style workloads. Whilst power is a critically important consideration, this is combined with the challenge that these workloads often operate within specific time constraints and as such an architecture that can deliver performance and power efficiency is critical.

It has already been demonstrated that reconfigurable architecture, such as field programmable gate arrays (FPGAs), can deliver IoT based solutions with significantly lower power consumption compared to central processing units (CPUs) or graphics processing units (GPUs), however a major restriction to leveraging FPGAs are the challenges in programming. An alternative is the use of soft-cores, where an FPGA is configured to appear and be programmed like traditional CPU core(s). This has resulted in the ability to specialise CPU architectures for the environment in question, in our focus the specific disaster which is of interest, without the impediment of expensive manufacturing costs associated with taping out physical chips. Furthermore, FPGA-based soft-cores enable researchers to experiment with CPU core designs, and this is important to understand the optimal design of an embedded micro-core architecture that will be deployed in the field.

When choosing an IoT architecture, whether it be a physical chip or soft-core, it is important that the choice made is a good one, however with over 40 implementations of the RISC-V architecture alone, the ability to assess competing designs simply and quickly is crucial. Furthermore, it is also important to consider not only different micro-core instruction set architectures (ISAs) but also variants within a particular processor ISA. This is especially true for RISC-V based designs due to the rich micro-architecture ecosystem. For instance, when selecting a RISC-V CPU there is a choice between many important aspects which all exhibit performance and power trade-offs. Without hard numbers from benchmarking to quantify the impact of such choices, it is difficult to make informed decisions.

Whilst it would appear that the best approach would be to run a set of the large number of currently available benchmarks on the micro-cores, the process isn't as simple as it would at first seem. There are a number of architectural features common to micro-cores that makes them significantly different from traditional CPUs and difficult to benchmark, including tiny amounts of on-chip RAM, and low-level knowledge specific to each device such as memory maps and linker files. Therefore, running existing benchmarks as-is on micro-cores is at best difficult and more than not often impossible. In order to be able to meaningfully compare and contrast different micro-core architectures, a benchmark framework is required to abstract over each architecture's idiosyncrasies and complexities. 

It is such a framework and exploration of common micro-core architecture that this paper describes. This text is organised as follows, in Section \ref{background} we describe the background and related work already done in the community around benchmarking soft-cores, before we describe the selected CPUs in Section \ref{hardware} and our Eithne framework and benchmarks in Section \ref{benchmarks}. We provide a comparison of benchmark results, including soft-core performance, power consumption, and code density in Section \ref{results}. We highlight some of the challenges posed by the immaturity of the RISC-V ecosystem in Section \ref{challenges} and then conclude in Section \ref{conclusions}, before discussing further work. 

\section{Background and related work}\label{background}
Whilst micro-core architectures have been around for for some time, the ability to perform a detailed exploration of the characteristics of the different options and how they compare and contrast against each other is somewhat limited. The experiments performed are often centred around very specific benchmarks or codes, and difficult to recreate on other technologies. This is typically even more pronounced for soft-cores, as there are often numerous configuration options to tune the architecture. For instance, in \cite{makni_comparison_2016} the authors detail a performance and resource comparison between Xilinx's MicroBlaze\cite{seely_microblaze_2017} and LEON3\cite{noauthor_leon3_nodate} to determine their suitability for multi-core embedded processors. These are both soft-cores and whilst they conclude that the performance between these technologies is similar, the survey omits to explore any of the configuration options provided by the MicroBlaze, which can have a significant impact on overall performance. 

By providing configurable logic blocks sitting within a sea of interconnect, FPGAs enable a chip to be programmed so that it operates electronically based upon a software design. By including other facets, such as fast memory and digital signal processor (DSP) slices within these chips, complex and high performance circuitry can be temporarily configured. Soft-cores are a software description of a CPU, which is then used by the tooling to configure the FPGA to represent this electronically. Whilst it is generally accepted that FPGAs are more power efficient than CPUs and GPUs\cite{kestur_blas_2010}, work done in \cite{castells-rufas_energy_nodate} concluded that this power efficiency extends to soft-core processors running on FPGAs. This is important in IoT use for disaster detection, as it potentially addresses the programmability challenges of FPGAs, whilst maintaining many of the power efficiency benefits of FPGAs.

A key feature of the existing body of work benchmarking micro-core architectures is that comparisons have been performed on a very piecemeal basis, with little thought in making the process reusable across other technologies, benchmarks or metrics. Therefore, a key question for the urgent computing community when considering whether this is an appropriate technology to deploy in the field, is firstly whether micro-cores do provide performance and power benefits, secondly whether soft-cores can provide improved power efficiency due to their use of FPGAs, and lastly if one was looking to choose such a technology then what characteristics would they be most interested in.



\section{Hardware and metrics surveyed in this paper}\label{hardware}
The key objective of this research has been to determine the impact of micro-core architecture features and configurations, such as pipeline depth, and hardware floating point support, on metrics of interest to use of these architectures for disaster detection. Based on this application, we are interested in the following metrics:

\begin{itemize}
  \item Performance, detailing the performance that the micro-core can deliver
  \item Power consumption, exploring the amount of power that the micro-core draws
  \item Energy cost, which combines performance and power consumption to consider the overall energy used for a workload
  \item Resource usage, which is for soft-cores only and explores the amount of resources that are required. This is important as increased resources result in larger, more expensive chips.
  \item Code density, which explores the memory size required for the specific codes being studied. This depends heavily on the ISA, and increased memory requirements result in increased cost and power consumption
  \item Ecosystem maturity, exploring the availability of and support for, the general software ecosystem on the technology in question.
\end{itemize}

These criteria directly influence the suitability of micro-cores for execution on the edge for disaster detection.
 
\subsection{CPUs}
In this paper various micro-core architectures, both physical chips and soft-cores, along with numerous configurations have been compared against to understand the benefits and negatives of deploying the technologies in this specific area.
The following micro-core CPUs were chosen for comparison, based on availability and, for soft-cores, ease of integration into an FPGA multi-core design:
\begin{itemize}
	\item PicoRV32 (soft-core)
	\item Xilinx MicroBlaze (soft-core)
	\item ARM Cortex-A9 (hard processor)
	\item Adapteva Epiphany-III (hard processor)
\end{itemize}

Whilst this list is a fraction of the available soft-cores, within the context of this paper only a limited set can be realistically benchmarked and we selected such a short-list due to their differences and interest. However, the Eithne benchmark framework described in Section \ref{eithne} is extensible and can can built upon to perform analysis of other technologies of interest.

The PicoRV32\cite{wolf_picorv32:_2018}  is a von Neumann, non-pipelined open source RISC-V soft-core that supports the RV32IMC instruction set\cite{waterman_design_2016}. This is the simplest soft-core considered in this paper, and the von Neumann architecture means that both code and data resides in the same memory space. This is at odds with many other micro-core technologies which are based on a Harvard architecture which allocate code and data in different memories. The Xilinx MicroBlaze is an example of this Harvard architecture approach, but furthermore it is pipelined. This means that, unlike the PicoRV32 which must wait until each instruction has entirely finished executing before the next can start, the processing of an instruction is split up across multiple stages. Each pipelined stage can execute different instructions, completing a specific subset before passing it onto the next. Whilst this makes the design more complex, increasing power consumption and resource usage, it means that multiple instructions are being processed concurrently which can result in performance benefits.

The PicoRV32 only provides hardware support for integer arithmetic, and floating point arithmetic must be implemented in software, typically provided by the underlying runtime. However, the Xilinx MicroBlaze contains an optional hardware floating point unit (FPU), and enables direct execution of hardware floating point arithmetic by the CPU. This increases performance, but at the cost of increasing complexity and potentially power usage, so a key question is whether such a trade-off is worth it given the typical workloads found in edge-based disaster detection and tracking.

The ARM Cortex-A9 and Adapteva Epiphany-III are physical micro-core CPUs. An important question is whether physical chips, which run at higher clock frequencies, exhibit significant benefits over their soft-core brethren for IoT workloads. The Cortex-A9 is a Harvard, superscalar, out of order, pipelined, architecture with FPU. Superscalar means that the architecture is capable of completing more than one instruction per clock cycle, and it typically achieves this by leveraging out or order execution, where the order of physical instruction execution is determined only by dependencies rather than the order in which they are issued by the programmer. This is a very common approach, and the CPU will commit completed instructions in program order to ensure a consistent view of execution on behalf of the programmers. The Epiphany-III is rather similar, with 16 von Neumann, superscalar cores, and 32KB of scratchpad static RAM (SRAM) per core. 

\subsection{Hardware boards}
The Xilinx PYNQ-Z2\cite{noauthor_pynq-z2_nodate} single board computer (SBC) was selected as the platform for all soft-core devices explored in this paper. The on-board Xilinx Zynq 7020 FPGA contains a dual-core 650MHz ARM Cortex-A9 running Linux, accessing 512MB of on-board but off-chip dynamic RAM (DRAM). The programmable logic of the Zynq 7020 contains 53,200 configurable LookUp Tables (LUTs), 4.9Mb of block RAM (BRAM) and 220 DSP slices which are commonly used for floating point arithmetic. Whilst one can configure a small number of cores with a large amount of memory per core, realistically eight cores, each with 32KB memory, is the best balanced that can be achieved, and that is the configuration we have selected. However, the reader is able to reuse our benchmarks and framework to explore the performance of other configurations. The Adapteva Parallella \cite{noauthor_parallella_2013} was used to host the Epiphany-III, providing a 650MHz dual-core ARM Cortex-A9 running Linux, with 1GB of DRAM and the 600 MB/s link to the Epiphany co-processor.


\section{Software benchmarks}\label{benchmarks}
Three benchmarks have been selected to compare the overall performance of the selected technologies. We have selected these benchmarks to test different facets of the technology, firstly the LINPACK single-precision \cite{dongarra_linpack_2003} benchmark has been chosen due to its role in stressing raw floating point performance, which will be interesting to explore in the context of the micro-cores. LINPACK determines the performance of a system in millions of floating point operations per second (MFLOPS) by performing LU factorization as follows \cite{linpack}

\begin{enumerate}
    \item Set up a random matrix $A$ of size $N$
    \item Set up the vector $X$ which has all values set to $1$
    \item Set up a vector $B$ which is the product of $A$ and the vector $X$
    \item Compute an LU factorization of $A$ 
    \item Use the LU factorization of $A$ to solve $A * X = B$
\end{enumerate}

The number of floating point operations required for the two LU factorizations is \[ops = 2 * N*N*N / 3 + 2 * N * N\] and the MFLOPS value is calculated by \[MFLOPS = ops / ( t * 1000000 )\]

In addition to LINPACK we have also implemented the Discrete Fourier Transform (DFT) and Fast Fourier Transform (FFT) benchmarks \cite{fftbenchmark}. These were chosen due to their relevance in the embedded community, and also represent a very common workload across numerous disaster scenarios, for instance interpreting streaming sensor data to identify any anomalies that might represent some sort of emergency, such as an earthquake. These Fourier benchmarks compute the forward and backwards transform of data. 

Whilst we have chosen these particular benchmarks to drive the exploration of micro-core characteristics in this paper, it is important to note that the Eithne benchmark framework as described in Section \ref{eithne} is easily extensible with other benchmarks that suit specific disaster use-cases. 

\subsection{Eithne Benchmark framework}\label{eithne}
To minimise the impact of different micro-core architectures on the benchmark results, the Eithne\footnote{Eithne (/\textipa{En{\textsuperscript{j}}9}/ "enya"):  Gaelic for "kernel" or "grain".} framework has been developed which enables a suite of benchmarks to be run across many different devices with limited or no modification required. The framework uses a single execution model across all devices, where the kernels are transferred to the device to be benchmarked, and a listener is launched awaiting data transfers and launch requests from the host benchmark application. This ensures that the communications link architecture, such as shared memory or universal asynchronous receiver-transmitter (UART), does not significantly impact the measurement of kernel execution. Data transfers are separated from kernel launch requests to enable the measurement of the communications link latency and bandwidth. Most importantly, this framework has been developed with extensibility in mind, where new devices, benchmarks, or metrics can be trivially added.
 
 \begin{figure}[h!]
	\centering
	\includegraphics[width=0.45\textwidth]{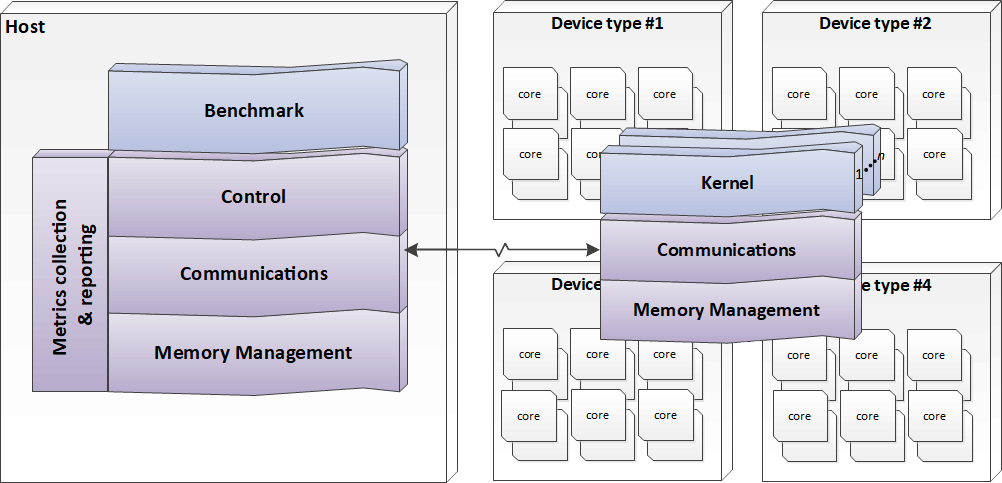}
	\caption{Eithne framework architecture}
	\label{fig:eithne}
\end{figure}

Figure \ref{fig:eithne} illustrates the Eithne framework's layered architecture, provided as a stack of functionality. Adding a new benchmark, 
device or communication mechanism only requires that specific layer of the framework is modified, with all other layers remaining unchanged. This not only simplifies supporting new technologies, benchmarks and metrics, but also isolates the remaining code, reducing the regression testing effort required.

\begin{figure}[h!]
	\centering
	\includegraphics[width=0.45\textwidth]{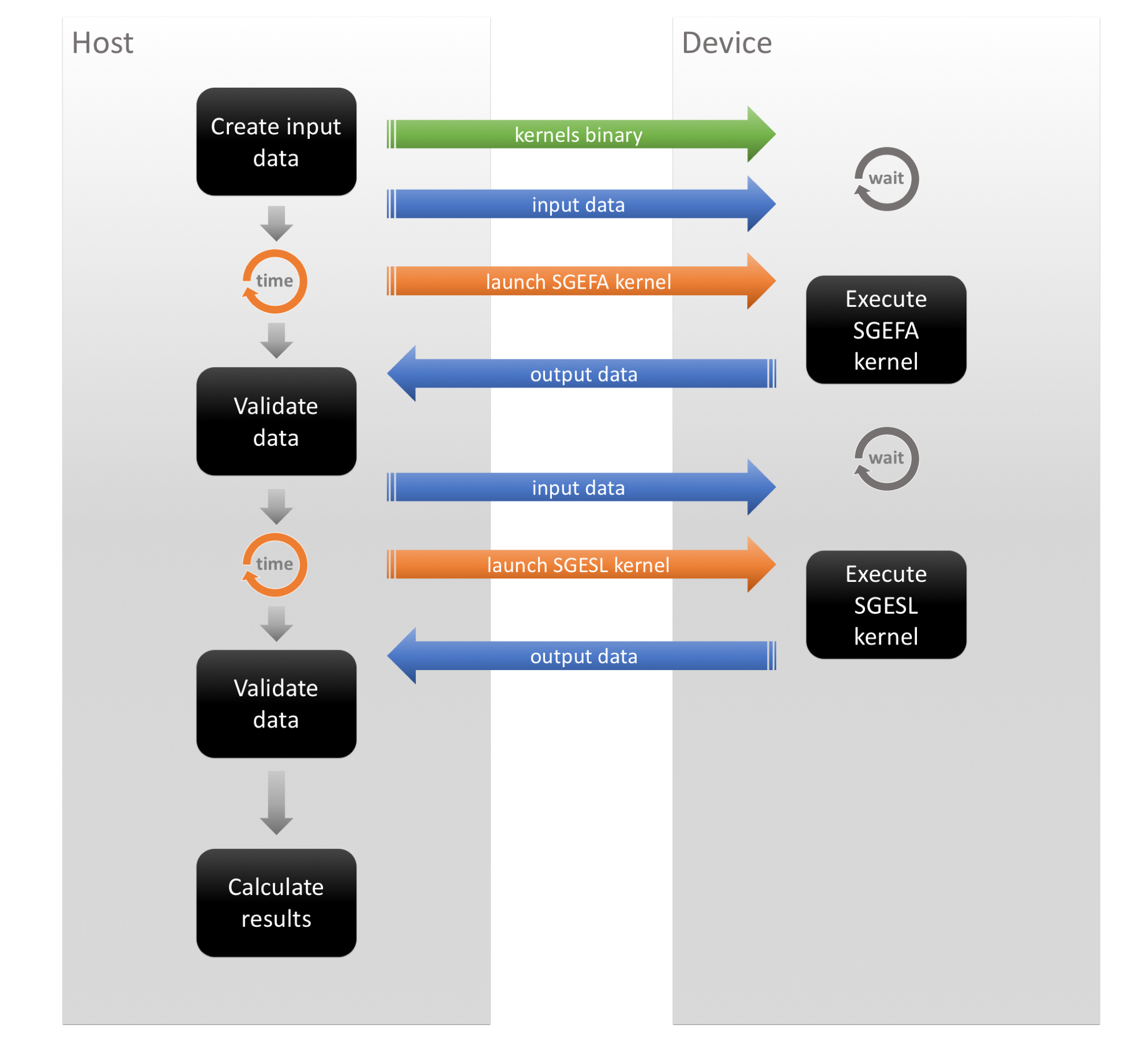}
	\caption{LINPACK benchmark host / device  flow}
	\label{fig:linpackflow}
\end{figure}

The high-level flow for the LINPACK benchmark is outlined in Figure \ref{fig:linpackflow} and in this paper we use LINPACK to outline the modifications required to run benchmarks on micro-core architectures using the Eithne framework. 

\subsection{LINPACK}\label{linpack}
A single-precision C version of the LINPACK\_BENCH benchmark \cite{linpack} was modified to run on the Eithne framework by separating out the \texttt{sgesl} and \texttt{sgefa} kernels, and their support functions, such as \texttt{saxpy} and \texttt{ddot}, from the rest of the initialisation, verification and timing code. These kernel codes were placed in a \texttt{kernels.c} file, along with the Eithne framework initialisation function which is outlined in Listing \ref{lst:kernelinit}. This code first registers the kernel input and output data variables;  \emph{a}, \emph{b}, \emph{ipvt}, \emph{job}, and \emph{info} in Listing \ref{lst:kernelinit}. These are then compiled and downloaded to the device, with kernel execution and data transfers being performed by the framework.

\begin{lstlisting}[style=cstyle, caption={Kernel framework initialisation function}, label={lst:kernelinit}]
void kernel_init(EithneTargetId id, EithneSharedMem buffer) {
  EithneKernel kernels[] = { sgefa, sgesl };

  EITHNE_INIT_DEVICE(vars,id,buffer+EITHNE_DATA_OFFSET,buffer,kernels);

  EITHNE_REGISTER_ARRAY(vars,A,EITHNE_FLOAT_ARRAY,a,N*LDA);
  EITHNE_REGISTER_ARRAY(vars,B,EITHNE_FLOAT_ARRAY,b,N);
  EITHNE_REGISTER_ARRAY(vars,IPVT,EITHNE_INTEGER_ARRAY,ipvt,N);
  EITHNE_REGISTER_SCALAR(vars,JOB,EITHNE_INTEGER,job);
  EITHNE_REGISTER_SCALAR(vars,INFO,EITHNE_INTEGER,info);

  EITHNE_START_LISTENER;  
}
\end{lstlisting}

The remaining LINPACK code was modified to use the Eithne framework API calls to allocate memory, register variables, transfer data and launch kernels. The host initialisation code for kernels running on the Adapteva Epiphany co-processor, Xilinx MicroBlaze and PicoRV32 soft-cores is outlined in Listing \ref{lst:hostinit}.

\begin{lstlisting}[style=cstyle, caption={Host framework initialisation code}, label={lst:hostinit}]
buffer = EITHNE_ALLOC_MEM(sizeof(float)*N*LDA);

EITHNE_INIT_HOST(vars,HOST_ID,buffer+EITHNE_DATA_OFFSET,buffer);
EITHNE_INIT_CORES(16);
EITHNE_START_CORES(16);

EITHNE_REGISTER_ARRAY(vars, A, EITHNE_FLOAT_ARRAY, a, N*LDA);
EITHNE_REGISTER_ARRAY(vars, B, EITHNE_FLOAT_ARRAY, b, N);
EITHNE_REGISTER_ARRAY(vars, IPVT, EITHNE_INTEGER_ARRAY, ipvt, N);
EITHNE_REGISTER_SCALAR(vars, JOB, EITHNE_INTEGER, job);
EITHNE_REGISTER_SCALAR(vars, INFO, EITHNE_INTEGER, info);
\end{lstlisting}

Listing \ref{lst:sgefa} outlines the addition of the required Eithne framework API calls to launch and time the \texttt{sgefa} kernel to the existing LINPACK code. The kernel function parameters are replaced by the \texttt{EITHNE\_SEND} and \texttt{EITHNE\_RECV} API calls. However, due to the previous registration API calls, the underlying kernel input and output variables such as \emph{a} and \emph{ipvt} are used unchanged, thereby minimising the impact to the existing codes.

\begin{lstlisting}[style=cstyle, caption={Executing and timing the LINPACK sgefa kernel}, label={lst:sgefa}]
/* Input to SGEFA */ 
EITHNE_SEND(vars, TARGET_ID, A); 

t1 = cpu_time ( );
EITHNE_EXECUTE(TARGET_ID, SGEFA);
t2 = cpu_time ( );

/* Output variables from SGEFA */ 
EITHNE_RECV(vars, TARGET_ID, A);
EITHNE_RECV(vars, TARGET_ID, IPVT);
EITHNE_RECV(vars, TARGET_ID, INFO);  
\end{lstlisting}

In all experiments \emph{N} was set to 20, this was found to be an appropriate parameter setting which takes into account the fact that the micro-cores have very limited memory. As the LINPACK matrix order size \emph{N} impacts the overall MFLOPS result, we felt it was important to keep this consistent across all devices to enable a like-for-like comparison of performance.

\subsection{DFT and FFT}\label{fft}
Both DFT and FFT benchmarks \cite{dft_fft_benchmark} were implemented using the Eithne framework. The main \texttt{dtf} and \texttt{fft} kernels and support functions were extracted from the surrounding initialisation and timing code and placed in a \texttt{kernels.c} file, along with the kernel framework code initialisation code similar to that outlined in Listing \ref{lst:kernelinit} but updated to reflect the FFT and DFT variables and kernels.

Eithne kernels have a void parameter list as the input and output variables are transferred by the framework. Since the \texttt{fft} kernel is recursive, a wrapper function was created to isolate the changes to the kernel code, as shown in Listing \ref{lst:fftkernel}.

\begin{lstlisting}[style=cstyle, caption={FFT kernel and wrapper}, label={lst:fftkernel}]
void fft_wrapper(void) {
  fft(xfer_sig, xfer_f, xfer_s, 1<<N, xfer_inv);
}

void fft(const Comp *sig,Comp *f,int s,int n,int inv) {
    int i, hn = n >> 1;
    Comp ep = comp_euler((inv ? PI : -PI)/(float)hn),ei;
    Comp *pi = &ei, *pp = &ep;
    if (!hn) *f = *sig;
    else
    {
        fft(sig, f, s << 1, hn, inv);
        fft(sig + s, f + hn, s << 1, hn, inv);
        pi->a = 1;
        pi->b = 0;
        for (i = 0; i < hn; i++)
        {
            Comp even = f[i], *pe = f + i, *po = pe + hn;
            comp_mul_self(po, pi);
            pe->a += po->a;
            pe->b += po->b;
            po->a = even.a - po->a;
            po->b = even.b - po->b;
            comp_mul_self(pi, pp);
        }
    }
} 
\end{lstlisting}

\section{Benchmark results}\label{results}
Based upon the software framework described in Section \ref{eithne}, the benchmarks of Section \ref{linpack} and Section \ref{fft} have been executed on the hardware described in Section \ref{hardware}. In this section we use these results as a tool to explore, and compare and contrast, the different characteristics of these technologies and consider their suitability for use on the edge in the context of urgent computing.

\subsection{Performance}
\label{sec:performance}
\subsubsection{LINPACK} 
\begin{figure}[h!]
	\centering
	\includegraphics[width=0.45\textwidth]{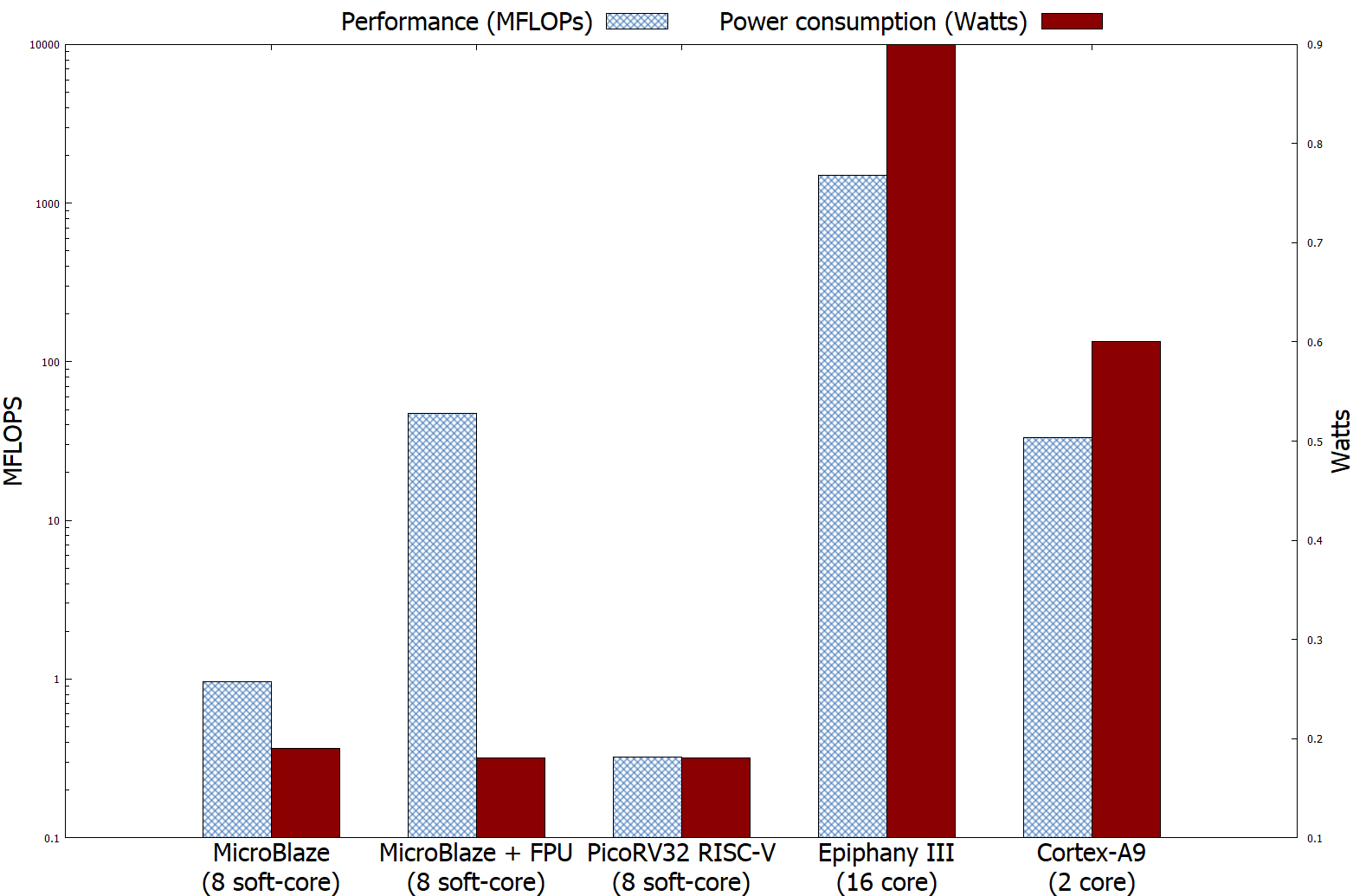}
	\caption{Micro-core LINPACK benchmark performance / power consumption}
	\label{fig:perfpower}
\end{figure}

Figure \ref{fig:perfpower} illustrates the performance of different micro-core technologies running the LINPACK benchmark via the Eithne framework. It can be seen that the physical micro-core processors have significantly higher performance than soft-cores, which can be mainly explained by the higher clock rates of between five and six times. However, for the Epiphany-III, there are other performance design features that expand this gap that will be covered in Section \ref{sec:dft_and_fft} below.

It is interesting to note that the power consumption for all soft-cores running on the Zynq 7020 is almost the same, even when the soft-cores have very different performance profiles, as is the case for the integer-only PicoRV32 and MicroBlaze with FPU. This initial analysis would seem to support the view that FPGA-based designs are more power efficient than physical CPUs, per \cite{kestur_blas_2010}. However, as we will discover in Section \ref{sec:energy_cost}, the overall power required to execute the code to completion is perhaps a more important figure influencing the choice of micro-cores deployed in the field.

\subsubsection{DFT and FFT}
\label{sec:dft_and_fft}
The performance results for the DFT and FFT benchmarks detailed in Table \ref{tbl:fftperf} are somewhat similar to those seen for LINPACK, where the Epiphany-III physical processor is 653 times faster than the PicoRV32. The integer-only MicroBlaze is 2.6 times faster than the PicoRV32 due to its pipelined architecture, and the hardware floating point MicroBlaze is 13.7 times faster than the PicoRV32. Assuming that the performance scaled linearly with clock frequency, the Epiphany would be 109 times faster than the PicoRV32, and 7.5 times faster than the MicroBlaze, when running at a clock speed of 100MHz.

\begin{table}[h!]
\caption{DFT / FFT Micro-core Performance}
\centering%
\footnotesize
\begin{tabular}{| c c c |}
\hline
\textbf{Device} & \textbf{DFT (seconds)} & \textbf{FFT (seconds)} \\
\hline
\textbf{PicoRV32} & 0.11096 & 0.11130 \\
\textbf{MicroBlaze} & 0.04259 & 0.04266 \\
\textbf{MicroBlaze \& FPU} & 0.00808	& 0.00825 \\ 
\textbf{Epiphany-III} & 0.00017 & 0.00017 \\
\hline
\end{tabular}
\label{tbl:fftperf}
\end{table}
 
When comparing the PicoRV32 against the MicroBlaze, one can observe the performance advantages of a pipelined architecture and hardware floating point support. However, results from the Epiphany-III highlight the further benefits of a 64 register, superscalar CPU that can execute two floating point and a 64-bit memory load operation every clock cycle \cite{noauthor_epiphany_2014}.

\subsection{Power consumption}
In Section \ref{sec:performance} it was seen that more complex architectures deliver significantly better performance than simpler ones, which is not a major surprise. Furthermore, the fact that physical processors can operate at a much higher clock frequency also delivers numerous performance benefits. However, a key question is whether there are any power consumption disadvantages of such designs, and this was calculated by measuring the voltage and amperage of each board running the benchmarks using a wiring harness and two UNI-T UT60E multimeters.

\subsubsection{LINPACK} 
Power consumption for the LINPACK benchmark, along with the measured floating point performance (in MFLOPS), is illustrated in Figure \ref{fig:perfpower}. From the results, it can be seen that the (integer only) MicroBlaze soft-core is five times more energy efficient than the Epiphany-III and 4 times more so than the Cortex-A9. The overall number of cores is likely to play a factor here, namely the fact that the Epiphany-III has 16 cores, the Cortex-A9 has 2, and the MicroBlaze was configured with 8 cores. 

When the clock rate (MicroBlaze 100Mhz, Epiphany 600MHz and Cortex-A9 650MHz) and the number of cores is taken into account, then we find for that each core, the Epiphany is approximately two times more power efficient than the MicroBlaze, but the MicroBlaze is six and a half times more power efficient than the Cortex-A9. By contrast, the PicoRV32 is about two times less power efficient than the MicroBlaze, and six times less power efficient than the Epiphany-III, but is still around two times more power efficient than the Cortex-A9. 

The fact that the much simpler PicoRV32 drew more power than the more complex MicroBlaze surprised us, but it can most likely be explained by the fact that more complex (AXI4) bus support logic is required for a multi-core PicoRV32 design, whereas the multi-core MicroBlaze uses a much simpler Local Memory Bus (LMB) design. 

As expected, power consumption increases with clock rate. However, as outlined, the Watts, MHz and core results for the Epiphany-III are twice as efficient than for the MicroBlaze, making it the most power efficient core design of those tested. This may be due in part to general power inefficiencies in the soft-core designs or the impact of the greater instruction decoding logic of the soft-cores versus the Epiphany-III. For instance, the MicroBlaze has 129 instructions and the Epiphany-III only has 41, and this is a consideration when one is running a micro-core in the field for this workload, namely based on the specialised nature is it possible to significantly limit the number of instructions?

\subsubsection{DFT / FFT} 
The larger DFT / FFT benchmark kernel binaries, due to the inclusion of \texttt{sin()} and \texttt{cos()} functions, required more BRAM than is available on the 8 soft-core devices. Therefore, 4 core designs for the MicroBlaze and PicoRV32 were synthesised, each with 128KB of scratch-pad memory. The MicroBlaze design also included hardware floating point support, MicroBlaze+FPU, implemented using FPGA DSP slices, unlike the integer-only MicroBlaze used in the 8-core bitstream for the LINPACK benchmark. 


It is interesting to compare the impact of enabling hardware floating point support in the MicroBlaze, and this impacted power consumption significantly, resulting in the simpler PicoRV32 drawing 14\% less power than the MicroBlaze+FPU running the DFT and FFT benchmarks.

\begin{table}[h!]
\caption{DFT and FFT Benchmark Power Consumption}
\centering%
\footnotesize
\begin{tabular}{| c c c |}
\hline
\textbf{Device} & \textbf{Idle (Watts)} & \textbf{Load (Watts)} \\
\hline
\textbf{PicoRV32} & 2.05 & 2.19 \\
\textbf{MicroBlaze} & 2.36 & 2.54 \\
\textbf{Epiphany-III} & 3.46 & 4.36 \\
\hline
\end{tabular}
\label{tbl:fftpower}
\end{table}

As detailed in Table \ref{tbl:fftpower}, the Epiphany-III uses approximately 1.8 times the power of both the PicoRV32 or MicroBlaze+FPU whilst running the benchmark. However, as for LINPACK, when we consider overall efficiency per core, we find that the Epiphany delivers a lower figure of 0.27 Watts/core at 600MHz against 0.55 Watts/core at 100MHz for the PicoRV32 and 0.63 Watts/core at 100MHz for the MicroBlaze+FPU. Bearing in mind that CPU power consumption increases with frequency \cite{datta_cpu_2014}, the Epiphany-III draws significantly less power than the soft-cores when scaled to the same clock rate of 100MHz, estimated at 0.045 Watts/core at 100MHz, a fourteen times greater power efficiency than the PicoRV32.

\subsection{Energy cost}
\label{sec:energy_cost}
Whilst the absolute power consumption of a processor is important, the \textit{power consumption to solution} is also of great interest. For instance, the power consumption required for processing streams of input data to generate a result which determines whether a disaster is unfolding or not. Effectively, such a metric describes the overall energy cost of the system, and power consumption to solution or energy (E) is defined as: \[E = Pt\]

\begin{figure}
	\centering
	\includegraphics[width=0.45\textwidth]{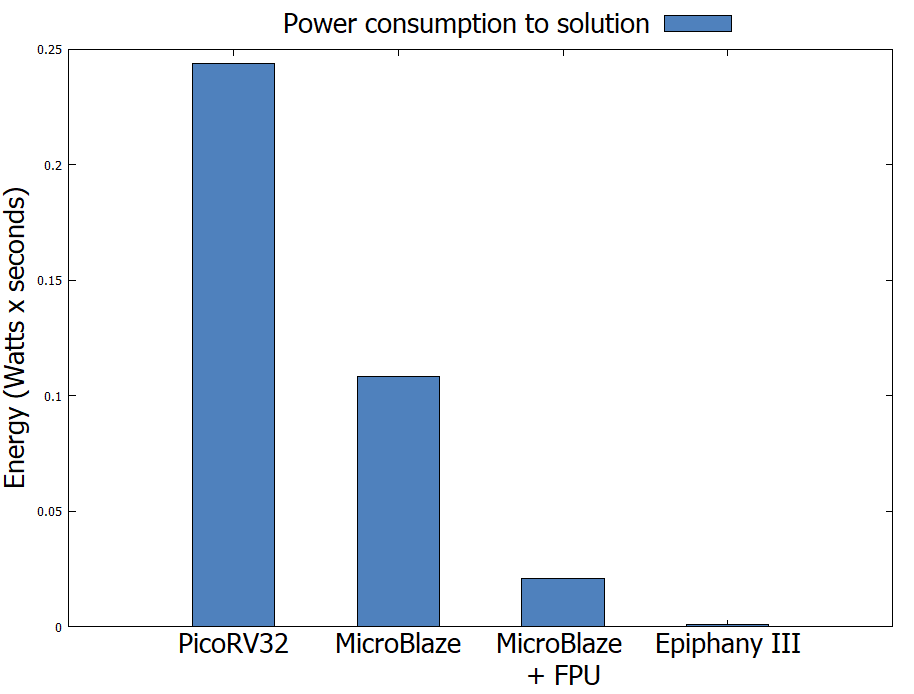}
	\caption{Micro-core FFT benchmark energy consumption}
	\label{fig:power2solution}
\end{figure}

Figure \ref{fig:power2solution} outlines the overall energy consumption for the FFT benchmark running on the selected micro-cores. The poor performance of the PicoRV32 results in a 328 times greater overall energy consumption than the much more powerful Epiphany-III processor for the same FFT kernel. The MicroBlaze+FPU, which is around fourteen times faster that the PicoRV32, uses around twelve times less energy overall to run the FFT benchmark than the PicoRV32. Therefore it can be seen here that the significant performance advantages of the MicroBlaze+FPU and Epiphany-III, mean that whilst the absolute power consumption is larger than the PicoRV32, there is still an overall energy benefit.

\subsection{FPGA area / resources}
\label{sec:usage}
Resource usage is important, as it dictates the overall size of the FPGA chip required to host specific soft-cores. Put simply, small FPGAs such as the Spartan are cheap \cite{alfke2009xilinx}, whereas as the number of resources scale up, the overall cost increases significantly. Table \ref{tbl:fpgaarea} details the FPGA resources utilised by each of the 8-core soft-core designs, where all of the soft-cores were configured to have a total of 64KB of on-chip RAM. This use of block RAM (BRAM) on-chip memory is the primary limiting factor in scaling the number of micro-cores further on the Zynq 7020. 

The integer-only MicroBlaze and PicoRV32 cores have a comparable LUT (look-up table) utilisation of 38\% and 35\% respectively. The PicoRV32 has been configured to support the RISC-V M (ISA MUL, DIV and REM instructions) and uses 15\% DSP slices for its implementation, whereas the MicroBlaze integer-only core does not use any DSP slices. The slightly increased LUTRAM and FF (flip-flop) requirements of the MicroBlaze over the PicoRV32 are likely to be attributable to the pipeline support and additional decoding logic required for the larger MicroBlaze instruction set versus the simpler RISC-V IMC instruction set.

The hardware floating point version of the MicroBlaze, MicroBlaze+FPU, uses 47\% more LUTs, 30\% more FF and 22 times more DSP slices than the integer-only MicroBlaze design. This represents a very significant increase in resources, and whilst the increase in DSP slice utilisation is to be expected, as this is the primary way in which floating point is executed by the FPGA, the increase in LUT usage was unexpected. Given an unlimited amount of on-chip memory, these figures would mean that the hardware floating point MicroBlaze could scale to 36 cores and the PicoRV32 to 53 cores on the Zynq 7020. 

\begin{table}[h!]
\caption{MicroBlaze and PicoRV32 Soft-core Z7020 Resource Percentage Utilisation}
\centering%
\footnotesize
\begin{tabular}{| c c c c c c |}
\hline
\textbf{Soft-core} & \textbf{LUT} & \textbf{LUTRAM} & \textbf{FF} & \textbf{BRAM} & \textbf{DSP} \\
\hline
\textbf{PicoRV32} & 35\% & 3\% & 13\% & 91\% & 15\% \\
\textbf{MicroBlaze} & 38\% & 6\% & 23\% & 91\% & 0\% \\
\textbf{MicroBlaze+FPU} & 56\% & 7\% & 30\% & 91\% & 22\% \\
\hline
\end{tabular}
\label{tbl:fpgaarea}
\end{table}

\subsection{Code density}
On-chip memory is often a major limiting factor when it comes to micro-cores. This is especially true with soft-cores, as from a resource usage perspective it was demonstrated in Section \ref{sec:usage} that BRAM is the factor that determines the overall number of soft-cores that can be implemented on a particular FPGA. Therefore, the size of the resultant kernel binaries is an important consideration with respect to the choice of processor instruction set architecture. 

All of the micro-cores selected for the benchmarks use the GNU C Compiler (GCC). Whilst it would be ideal to use the same version of GCC across all devices, some micro-cores, such as the Epiphany-III and RISC-V, only support specific versions of GCC. Therefore, for the benchmarks, we used the recommended version of GCC for each micro-core architecture. Apart from micro-core specific linker files and compiler flags (little-endian for the MicroBlaze), the GCC compiler options were identical across all architectures. Due to the aforementioned RAM limitations, the kernel codes were optimised for size (GCC option \texttt{-Os}), rather than for speed (GCC option \texttt{-O3}).

\begin{figure}[h!]
	\centering
	\includegraphics[width=0.45\textwidth]{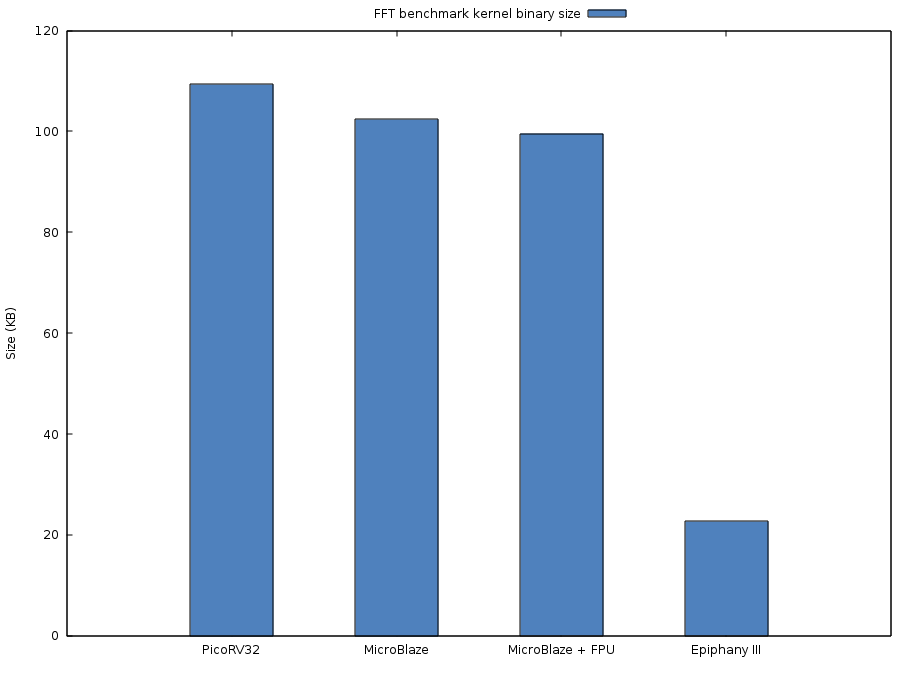}
	\caption{Micro-core FFT benchmark kernel size}
	\label{fig:kernelsize}
\end{figure}

Figure \ref{fig:kernelsize} illustrates the FFT kernel binary size produced by GCC for the micro-cores. The three soft-core options are roughly equivalent at 109KB for the PicoRV, 102KB for the integer-only MicroBlaze and 100KB for the hardware floating point MicroBlaze. The kernel binary for the MicroBlaze with hardware floating point is slightly smaller than the binary for the integer only core, as that must include software floating point emulation routines. However, at 23KB, the FFT kernel binary size for the Epiphany is significantly smaller than for the soft-cores. 

This difference is stark, and a surprising aspect of this comparison is that a larger instruction set does not seem to result in a smaller binary file size. For example, the RV32IMC ISA has 76 instructions, the MicroBlaze ISA 129 and the Epiphany-III ISA 41 instructions. Therefore, one would assume that the Epiphany-III binary would be the largest, as fewer instructions must be explicitly composed together. Yet the Epiphany has, by far, the smallest binary file size. 

In fact, this supports the assertion in \cite{mutigweInstructionSetUsage2013} where the authors state that one does not need more than 100 instructions, and further research in  \cite{mutigweInstructionSetUsage2013} also demonstrates that GCC only uses between 5-20\% of all the instructions across a range of processors (x86, x86-64, PowerPC and MIPS). When one considers the silicon area and power requirements of a CPU's instruction set decoding logic, it would seem prudent to keep a micro-core's ISA as small as possible. This could be a key factor in the Epiphany-III's impressive performance and power consumption figures that we obtained for the LINPACK, DFT and FFT benchmarks.

\subsection{Maturity of the software ecosystem}\label{challenges}
The Epiphany-III and MicroBlaze are commercial products and, the MicroBlaze especially which is supported by Xilinx, offer a fairly large software ecosystem. This is especially the case for C compilation, which is very popular in embedded computing. Furthermore, both these technologies can be obtained \emph{off the shelf}, and simple to operate. However, whilst there are over forty RISC-V based implementations available, for this paper, we have found that actually being able to configure multi-core FPGA designs with these is a different matter. There are a number of challenges to successfully using RISC-V soft-cores, ranging from the ability to synthesise the source files, often designed for simulation rather than for use as an IP component within an FPGA bitstream, to the immaturity of the development tools. 

During the work conducted in this paper, we have observed a number of idiosyncrasies of the RISC-V ecosystem that we found especially challenging.

\subsubsection{Lack of RISC-V soft-core verification} 
Many of the available open source RISC-V soft-cores have not been verified against the published RISC-V ISA standards. For example, the VectorBlox ORCA \cite{noauthor_risc-v_2019} is stated \cite{risc-v-list_2019} to support the RV32IM ISA and provides options for hardware multiply and divide instructions, but does not implement the REM instruction specified in the M ISA option. The  consequence is that that codes compiled by the official RISC-V GCC toolchain with this option enabled will freeze on the ORCA, and resulted in it being excluded from our comparisons.  Tracking down this type of issue is time consuming and beyond the expertise of many programmers due to the lack of support for debugging tools on a number of the available soft-cores, such as the PicoRV32 and ORCA. 

\subsubsection{Low-level GCC linker file knowledge}
The RISC-V uses register \texttt{x2} as the stack pointer, which needs to be initialised before C functions are called. This is usually performed by an assembly language routine called when the binary is loaded onto the core. For simple codes, this initialisation routine can be compiled or linked without issue. However, for more complex codes requiring the \texttt{-ffunction-sections -Wl,-gc-sections} GCC optimisation flags to reduce the size of the binary, the GCC optimiser will remove the initialisation code because it is managed in the linker file and the code will fail to run on the core. Ensuring that GCC does not remove this code but in a manner where the compiler can still perform size optimisation requires in-depth understanding of segments and modification of the RISC-V linker file. Therefore, compiling and running codes such as the LINPACK and FFT benchmarks on RISC-V soft-cores is far more involved than simply taking existing codes and recompiling them for the RISC-V. 

\subsubsection{Inconsistent RISC-V compressed instruction set support}
Although RISC-V supports a compressed instruction set (RV32C) that can reduce the size of a compiled binary by up to 30\% \cite{patterson_risc-v_2017}, the majority of the 32 bit RISC-V soft-cores do not support compressed instructions. When we consider that the FFT kernel binaries compiled using the RV32IMC ISA were too large to fit in 64KB, and as such we had to reduce our design down to four 128KB cores, the provision of compressed instruction support could provide significant benefits for code density.

\section{Conclusions and further work}\label{conclusions}
Micro-core architectures have significant potential in the monitoring of the environment, operating on streams of sensor data to detect disasters. Running on the edge, important characteristics of the technology differ greatly from those typically found in the data-centre. However, given the choice of architectures available, an important question is which one is most applicable, the answer to which will often depend upon the situation in question.

In this paper, to address this, we have introduced a framework that greatly simplifies the benchmarking of micro-cores. Layering our design makes it far easier to plugin new architectures, benchmarks, and metrics, which until now has been very difficult to achieve. We then ran a series of benchmarks across some of the more popular micro-core architectures, exploring resulting measurements based upon characteristics most important in the IoT domain. 

We demonstrated that, irrespective of clock frequency, the Epiphany-III physical chip delivers significant performance and is more energy efficient than the other architectures. Given FPGA processor implementations have between eighteen and twenty six times greater circuit delay than their custom CMOS \cite{wong_comparing_2011} equivalents, this is not surprising, however it does illustrate that performance is important when optimising for energy efficiency. However, in terms of absolute power draw the PicoRV32 consumed around half the power of the Epiphany-III. We have also shown that the most important limiting factor for soft-cores tends to be the amount of on-chip memory, BRAM, present. Therefore, the size of kernel binaries produced for a particular processor's ISA is a critical limitation to the number of viable cores that can be implemented. 

Whilst \cite{castells-rufas_energy_nodate} found that their many soft-core processor was highly energy efficient, our results paint a different picture. They estimated the energy efficiency of their customised soft-core was 1623 MOPS/Watt, compared to 26 MOPS/Watt for an i7 running eight threads. However, we measured the MicroBlaze soft-core at 6 MFLOPS/Watt and the ARM Cortex-A9 at 55 MFLOPS/Watt. There is a difference here, for instance \cite{castells-rufas_energy_nodate} measured operations per second, and us explore floating point operations per second, but it is demonstrated by this work that the power efficiency of physical processors is at least nine times higher than the soft-cores. 

Further work includes extending our benchmarking framework to include other micro-core architectures, and to explore other relevant codes to disaster detection. Specifically, we think that connecting to real-world sensors and supporting the streaming in of data would be a sensible next benchmark. This would be another possible metric, how fast data can be streamed into a chip, and one where the soft-cores might have an edge due to the large number of IO connections that some FPGAs possess. Eithne currently separates data communications / transfers from the execution of kernels, therefore it has the support to enable the measurement of data bandwidth. Furthermore, there are embedded GPUs, such as NVIDIA's Jetson that would be interesting to also compare against. In terms of the micro-core architectures selected, there are higher performance RISC-V implementations, and exploring some of the customisable CPUs developed by SiFive would also be of interest.

Therefore, we conclude that micro-core architectures have potential for use in disaster detection, however this is best done with physical chips rather than soft-cores. Our hypothesis that soft-cores could provide the best of all worlds; high performance, energy efficiency and programmability is simply not the case. For production edge computing then one should utilise physical chips, such as the Epiphany-III, with soft-cores useful for prototyping and the exploration of architectural configurations.

\bibliographystyle{./bibliography/IEEEtran}
\bibliography{./bibliography/IEEEexample}

\end{document}